\documentclass{article}
\usepackage{spconf,amsmath,graphicx}
\pdfoutput=1
\usepackage{amsmath,amsfonts,amssymb}
\usepackage{diagbox}
\usepackage{float}
\usepackage{adjustbox}
\usepackage{graphicx}
\usepackage{enumitem}
\usepackage{algorithm}
\usepackage{amsfonts,amssymb}
\usepackage{tikz}
\newcommand{\LArrow}[1]{%
\parbox{#1}{\tikz{\draw[->](0,0)--(#1,0);}}
}

\usepackage{algpseudocode}

\setlist{nosep, leftmargin=14pt}

\usepackage{mwe} % to get dummy images

\title{Structural Constrained Virtual Histology Staining For Human Coronary Imaging using Deep Learning}
\name{Xueshen Li$^{1}$, Hongshan Liu$^{1}$, Xiaoyu Song$^{3}$, Brigitta C. Brott$^{2}$, Silvio H. Litovsky$^{2}$, Yu Gan$^{1}$}
\address{$^{1}$Department of Biomedical Engineering, Stevens Insitute of Technology, Hoboken, USA\\
$^{2}$School of Medicine, The University of Alabama at Birmingham, Birmingham,USA \\
$^{3}$The Icahn School of Medicine at Mount Sinai, New York, USA }

\begin{document}
%\ninept
%
\maketitle
%
%\begin{abstract}
%Magnetically controlled capsule endoscope (MCCE) is an emerging tool for diagnosis of gastric diseases. The performance of MCCE is similar to that of transoral gastroscopy while possessing advantages of comfort, safety, and no anesthesia. In this project, we propose a deep learning algorithm to detect human gastric peristalsis (contraction wave) using MCCE. Based on convolutional neural network (CNN) and long short-term memory (LSTM), we achieve 87.3\% accuracy, 77.6\% sensitivity, 86.8\% precision, and 87.9\% specificity on 38 MCCE data records for detecting gastric contraction waves. Our method has great potential in assisting diagnosis of gastric diseases by evaluating gastric mobility.
%\end{abstract}
\begin{abstract}
Histopathological analysis is crucial in artery characterization for coronary artery disease (CAD). However, histology requires an invasive and time-consuming process. In this paper, we propose to generate virtual histology staining using Optical Coherence Tomography (OCT) images to enable real-time histological visualization. We develop a deep learning network, namely Coronary-GAN, to transfer coronary OCT images to virtual histology images. With a special consideration on the structural constraints in coronary OCT images, our method achieves better image generation performance than the conventional GAN-based method. The experimental results indicate that Coronary-GAN generates virtual histology images that are similar to real histology images, revealing the human coronary layers.
\end{abstract}
\begin{keywords}
Virtual histology, Coronary artery disease, Optical coherence tomography, Deep learning
\end{keywords}
\section{Introduction}
Coronary artery disease (CAD) is the third leading cause of mortality worldwide and is associated with 17.8 million deaths annually \cite{hajar2017risk}. Apart from mortality, CAD can lead to symptoms involving angina, shortness of breath, heart attack, heart failure, and depression \cite{Lesperance.2000}. The morbidity and implications of CAD emphasize the importance of appropriate diagnosis and management. During percutaneous coronary intervention (PCI), a procedure to treat CAD, a stent is placed to open and keep open narrowed arteries. 
%Providing better interventions, with the capability to characterize coronary arteries, to improve the care of cardiovascular disease is an unmet need nationwide.
There is a significant need to improve coronary imaging to guide cardiac care.
%is the leading cause of cardiovascular mortality worldwide, with more
% than 4.5 million deaths occurring in the developing world \cite{Okrainec.2004}, and 1 in 7 deaths in the
% U.S \cite{EmeliaJ.Benjamin.2019}. Apart from fatality, CAD can potentially lead to symptoms involving angina, shortness of breath, nausea, heart attack, and depression \cite{Lesperance.2000}. The morbidity, fatality, and implications of CAD emphasize the importance of diagnosis and precautions. 

One valuable direction is to add histopathological visualization on real-time OCT imaging. Currently, histopathological analysis relies upon an off-line evaluation that requires postmortem evaluation. As an invasive and time-consuming process, histopathology involves fixation of tissues and preparation of thin stained sections followed by microscopic imaging. Multiple reagents are required to process histopathology and may introduce irreversible effects on tissue imaging. The process commonly takes a few hours to several days. It is thus not available for real-time analysis for tissue characterization of coronary arteries during PCI.

%\textbf{TODO: Why histology analysis is important.} Histological haematoxylin and eosin–stained (H\&E) tissue sections are preferred for histology analysis of coronary samples since it provides precise morphological information on various tissue types \cite{Goss2009TheoryAP}. However, tissue sectioning is required for histology analysis, which is invasive. Moreover, histology staining steps are time-consuming. For example, it takes hours to days to perform an H\&E staining task, involving protocols such as dewaxing, dehydration, hematoxylin, differentiation, bluing, eosin, dehydration, cearing, and
%cover‐slipping. On the other hand, 

Optical coherence tomography (OCT) is capable of resolving coronary tissue structures \cite{Tearney.2012}. It has been considered an optimal imaging system to assess plaques prior to stenting to ensure successfully stent deployment, and to assess vascular response to interventions \cite{wijns2015optical,jones2018angiography}. However, real-time interpretation of OCT images requires extensive training and prior knowledge and is not at the detailed analytic level as histopathological analysis.
\begin{figure}[t]
\centering
\includegraphics[width=8.5cm]{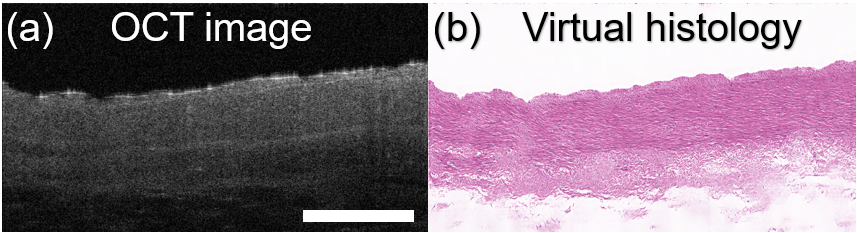}
\caption{(a) An example OCT image of a human coronary sample. The scale bar represents 1mm. (b) The virtual histology images that are generated from the example OCT image in (a). The virtual staining process can be done in real-time. The virtual histology image can help in interpreting OCT images.}
\label{fig:VirtualHistology}
\vspace{-15pt}
\end{figure}

Therefore, there is a movement to generate virtual histology images from OCT images \cite{Winetraub.2021}. An example of generating virtual histology images from OCT images for human coronary is shown in Figure \ref{fig:VirtualHistology}. Generative Adversarial Networks (GANs), e.g., Pix2Pix GAN, have been developed for virtual histology generation. However, such method requires a pixel-wisely paired OCT and H\&E image dataset. Generating the pixel-wisely paired dataset requires a substantial amount of effort, including encasing samples in fluorescent gel, photo-belching, and manually fine alignment, during pre-processing, imaging, and post-processing of the samples \cite{Winetraub.2021}. Moreover, the Pix2Pix GAN is trained from the pixel-wisely paired dataset, without considering high-level structural information. It is thus not optimized for virtual staining biological samples which has unique structural patterns. 

In this paper, we develop a deep learning network, namely Coronary-GAN, to generate virtual histology images from human coronary OCT images. Our Coronary-GAN is optimized for human coronary samples by incorporating the structural information of the coronary. In summary, the proposed method has the following contributions:

(1) We proposed a generative method to produce colorful virtual stain coronary structures from gray-scale human OCT images using deep learning.

(2) We proposed a Coronary-GAN, which does not require a pixel-wisely paired dataset of OCT and H\&E images. The proposed method significantly alleviates the effort for generating the dataset in comparison with existing methods.  
\begin{figure}[t]
\centering
\includegraphics[width=8cm]{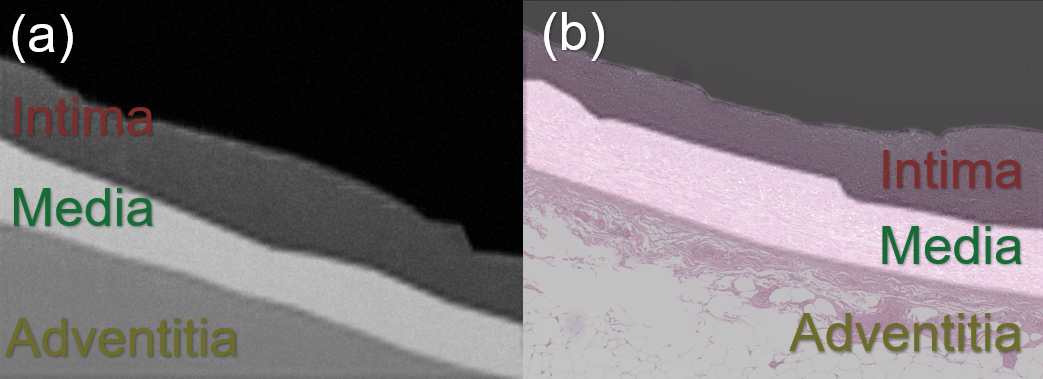}
\caption{Representative examples of (a) a labeled OCT image and (b) a labeled H\&E images. The intima, media, and adventitia are labeled. The OCT image and H\&E image are not pixel-wisely paired.}
\label{fig:labelOCTHE}
\vspace{-10pt}
\end{figure}

\begin{figure*}[h]
\centering
\includegraphics[width=17cm]{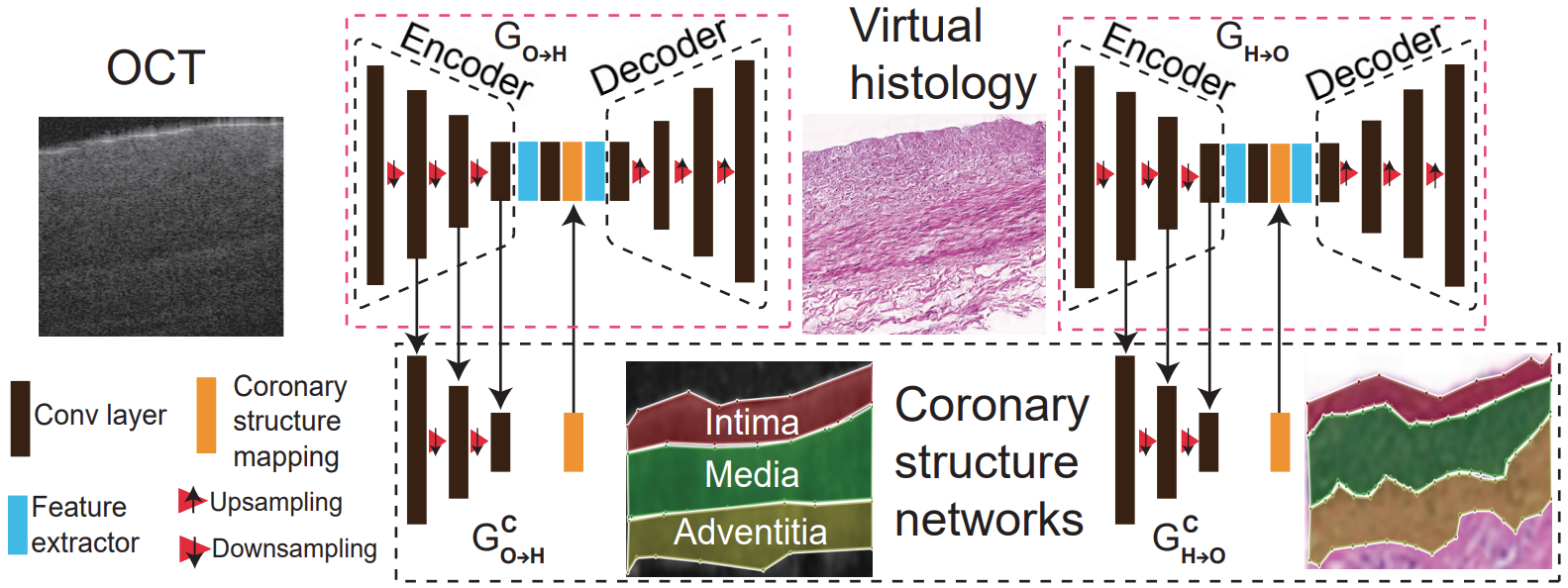}
%\caption{Scheme of the proposed Coronary-GAN. The Coronary-GAN consists of two generators, G$_{O\LArrow{.15cm}H}$ and G$_{H\LArrow{.15cm}O}$. Each generator has an encoder-decoder structure. The Coronary-GAN contains a coronary structure network, which segments the three layer structure of human coronary samples.}
\caption{Scheme of the proposed Coronary-GAN. The Coronary-GAN consists of two generators. Each generator has an encoder-decoder structure. The Coronary-GAN contains a coronary structure network, which segments the three layer structure of human coronary samples.}
\label{fig:CoronaryGAN}
\vspace{-10pt}
\end{figure*}
\section{Method}
\subsection{OCT and H\&E data acquisition}
Human coronary samples were collected from the School of Medicine at the University of Alabama at Birmingham (UAB). Specimens were imaged via a commercial OCT system (Thorlabs Ganymede, Newton, NJ) \cite{9779525}. A total of 285 OCT volumes were collected from 23 specimens with an imaging depth of 2.56 mm. The width of the images ranged from 2 mm to 4 mm depending on the actual sample size. Among all the volumes, the pixel size was 2 µm × 2 µm within a B-scan. Samples were processed for Hematoxylin and Eosin (H\&E) histology after imaging.

\subsection{Generation of human coronary structural database}
The human coronary consists of three layers, namely intima, media, and adventitia (from the inner part to the outer part). We labeled the three-layer structure in both OCT and H\&E images. An example of labeled OCT and H\&E images are shown in Figure \ref{fig:labelOCTHE}.

\subsection{Design of Coronary-GAN}
\subsubsection{Network architecture}
Details of the Coronary-GAN are shown in Figure \ref{fig:CoronaryGAN}. The Coronary-GAN consists of a Cycle-GAN network and a coronary structure network. G$_{O\LArrow{.15cm}H}$ transfers images from OCT domain to histology domain; G$_{H\LArrow{.15cm}O}$ transfers images from the histology domain to OCT domain.
The coronary structure network ensures the structural information of samples can be encoded and decoded by G$_{O\LArrow{.15cm}H}$ and G$_{H\LArrow{.15cm}O}$ respectively. Thus, the virtual staining network is optimized for human coronary samples by learning coronary structural information.
The Coronary-GAN does not require a pixel-wisely labeled OCT and H\&E dataset.
\subsubsection{Loss function}
The loss function $L$ of Coronary-GAN consists of four types, which are adversarial loss $L_{adv}$, cycle-consistency loss $L_{cycle}$, embedding loss $L_{embedding}$, and coronary structural loss $L_{coronary}$. 
\begin{equation} \label{eq1}
\begin{split}
&L(G_{O\LArrow{.15cm}H}, G_{H\LArrow{.15cm}O}, D_H, D_O, G^{C}_{O\LArrow{.15cm}H}, G^{C}_{H\LArrow{.15cm}O}) 
\\
& = L_{adv}(G_{O\LArrow{.15cm}H}, D_H) +  L_{adv}(G_{H\LArrow{.15cm}O}, D_O)
\\
&+ \alpha L_{cycle}(G_{O\LArrow{.15cm}H}, G_{H\LArrow{.15cm}O}) + \beta L_{embedding}(G_{O\LArrow{.15cm}H}, G_{H\LArrow{.15cm}O}) \\
&+ \gamma L_{coronary}(G^{C}_{O\LArrow{.15cm}H}, G^{C}_{H\LArrow{.15cm}O})
\end{split}
\end{equation}
The $L_{adv}$ and $L_{cycle}$ are described in \cite{8237506}, while $\alpha$, $\beta$, and $\gamma$ are hyper-parameters. Symbols $O$ and $H$ stand for OCT and Histology images respectively. G$_{O\LArrow{.15cm}H}$ and G$_{H\LArrow{.15cm}O}$ are two generators that generate virtual histology images from OCT images and virtual OCT images from histology images respectively. D$_H$ is the discriminator for histology images and D$_O$ is the discriminator for OCT images. G$^{C}_{O\LArrow{.15cm}H}$ and G$^{C}_{H\LArrow{.15cm}O}$ are the coronary structure networks.

We enforce the structural constraint of human coronary samples by developing an embedding loss function and a coronary structural loss function. In our design, the encoders G$^{en}_{O\LArrow{.15cm}H}$ and G$^{en}_{H\LArrow{.15cm}O}$ extract the embedding information of the coronary samples from corresponding OCT and H\&E images. Then the decoders G$^{de}_{O\LArrow{.15cm}H}$ and G$^{de}_{H\LArrow{.15cm}O}$ perform virtual staining and virtual OCT scanning. We design the $L_{embedding}$ term, minimizing the differences between the embeddings of two generators using L1-norm. 

\begin{equation} \label{eq2}
\begin{split}
L_{embedding} &= \mathbb{E}_H[\|G^{en}_{H\LArrow{.15cm}O}(G_{O\LArrow{.15cm}H}(O))-G_{O\LArrow{.15cm}H}(O)\|_1]
\\
& + \mathbb{E}_O[\|G^{en}_{O\LArrow{.15cm}H}(G_{H\LArrow{.15cm}O}(H))-G_{H\LArrow{.15cm}O}(H)\|_1]
\end{split}
\end{equation}
G$^{C}_{O\LArrow{.15cm}H}$ and G$^{C}_{H\LArrow{.15cm}O}$ predict the pixel-wise segmentation of the feature map acquired by coronary structure mapping of embeddings of encoders. The training of coronary structure networks is supervised by the human coronary structural database using cross-entropy loss terms.
\begin{equation} \label{eq2}
\begin{split}
L_{coronary} =& \mathbb{E}_H[-N_H^{-1}\sum^{N_H}_{n=1}\sum^{C}_{c=1}y^{n,c}_H\log(G^{C}_{O\LArrow{.15cm}H}(O))] \\
& +\mathbb{E}_O[-N_O^{-1}\sum^{N_O}_{n=1}\sum^{C}_{c=1}y^{n,c}_O\log(G^{C}_{H\LArrow{.15cm}O}(H))]
\end{split}
\end{equation}
$N_H$ and $N_O$ stand for the number of pixels in the embeddings acquired from histology and OCT images respectively. $y_H$ and $y_O$ are the labels from the human coronary database for histology and OCT images respectively. $C$ stands for the number of categories of the coronary layers (c=3). We aim to solve the following minmax optimization problem:
\begin{equation} \label{eq2}
\begin{split}
&G^{*}_{O\LArrow{.15cm}H}, G^{*}_{H\LArrow{.15cm}O}=
\\
& \arg \min\max L(G_{O\LArrow{.15cm}H}, G_{H\LArrow{.15cm}O}, D_H, D_O, G^{C}_{O\LArrow{.15cm}H}, G^{C}_{H\LArrow{.15cm}O})
\end{split}
\end{equation}

\section{EXPERIMENTAL DESIGN AND RESULTS}

\subsection{Experimental dataset}
From the 285 OCT volumes, we use 104 OCT images and 104 H\&E images that reveal intima, media, and adventitia for training and testing. We randomized 52 OCT images and 52 H\&E images for training. The other images are used as testing set. The human coronary database were created for the training purpose.
The OCT and H\&E images in the training set are divided into non-overlap patches with a size of 288$\times$288. We randomly flip the patches from left to right for data augmentation. The training set contains 1926 OCT image patches and 1926 H\&E image patches.

\subsection{Network implementation}
We implement the Coronary-GAN (shown in Figure \ref{fig:CoronaryGAN}) using PyTorch. For the encoders, we adopt 3 convolution layers with a stride of 2; for decoders, we adopt 3 transpose convolution layers also with a stride of 2. The feature extractor consists of 5 residual blocks as described in \cite{He2016DeepRL}. We use a convolution layer with a stride of 1 to perform coronary structure mapping. The G$_{O\LArrow{.15cm}H}$ takes 1 color channel for input and outputs 3 color channels; the G$_{H\LArrow{.15cm}O}$ takes 3 color channels and outputs 1 color channel. We follow the discriminator design in \cite{8237506}.

\begin{figure*}[h]
\centering
\includegraphics[width=17cm]{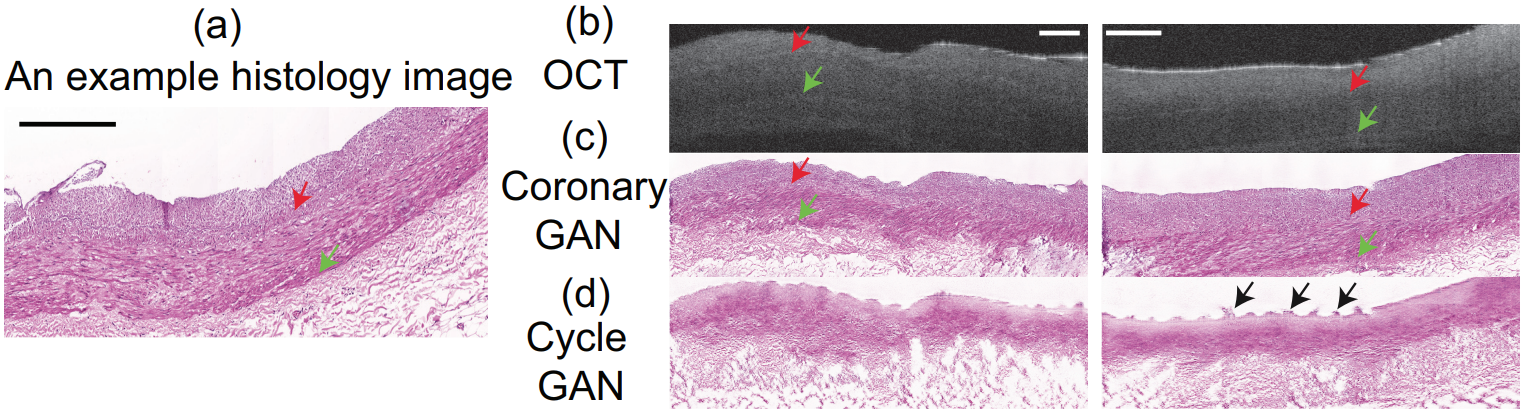}
\caption{The visual inspection of two representative OCT images and corresponding virtual histology images. (a) An example H\&E stained histology image for reference. (b) Representative OCT images. (c) Virtual histology images generated by Coronary-GAN. (d) Virtual histology images generated Cycle-GAN. The red arrows indicate the interface between intima and media; the green arrows indicate the interface between intima and adventitia. The black arrows indicate artifacts generated by Cycle-GAN. The scale bars represent 200$\mu m$.}
\label{fig:CoronaryGAN_result}
\vspace{-15pt}
\end{figure*}

\subsection{Evaluation setup and metrics}
\subsubsection{Network training details}
The pixel values of OCT and H\&E images were scaled to [0, 1]. The batch size was 16. The learning rate was initialized as $10^{-4}$, followed by a linearly decaying decay for every 2 epochs. $\alpha$, $\beta$, and $\gamma$ were set to 10, 5, and 5 respectively. In total, the networks were trained 10,000 epochs to ensure convergence. The experiments were carried out in parallel on 2 RTX A6000 GPUs. 
\subsubsection{Quantitative evaluation metrics}
We measure the similarity \textcolor{black}{from pairs of virtual histology and real histology images} using Perceptual Hash Value (PHV). This metric has been used in evaluating the quality of virtual histology images \cite{Liu.2021}.
\begin{equation}
PHV=\frac{1}{N}\sum^{n=1}_{N} H[|avg(F_i(\hat{H}))_{n} - avg(F_i(H))_{n}| - T]
\end{equation}
The PHV takes advantage of a pre-trained Resnet-101 network. $H$ and $\hat{H}$ represent the real histology and virtual histology images. $F_i(\cdot)$ stands for the feature maps that are generated from $i_{th}$ layer of the Resnet. N is the total channel number of extracted features after pooling. $avg(\cdot)_{n}$ stands for $n^{th}$ channel after average pooling operation. $H[\cdot]$ is the unit step function and T is a predefined threshold. 
%\textbf{Perceptual Hash Value.}
\subsection{Results}
\begin{figure}[t]
\centering
\includegraphics[width=8.5cm]{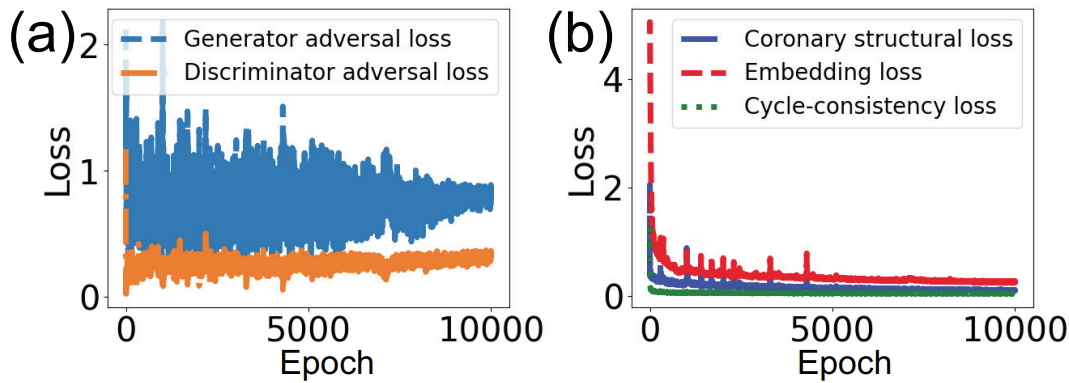}
\caption{Plot of loss functions during training of Coronary-GAN over 10,000 epochs. (a) Adversarial loss for generators and discriminators. (b) Cycle-consistency loss, embedding loss, and coronary structural loss.}
\label{fig:loss}
\end{figure}

\subsubsection{Network convergence}
We trained the Coronary-GAN for 10,000 epochs. The plots of loss functions are shown in Figure \ref{fig:loss}. Within 10,000 epochs of training, the Coronary-GAN converges for each loss term.
\subsubsection{Coronary-GAN provides better quantitative scores than Cycle-GAN}
As suggested in \cite{Liu.2021}, we observe the PHV scores calculated using $i=1$, $i=2$, and $i=3$. 
%We calculated the PHV scores using the feature maps extracted from layer1 (PHV$\_1$), layer2 (PHV$\_2$), and layer3 (PHV$\_3$). 
The resulting PHV scores are named as PHV$\_1$, PHV$\_2$, and PHV$\_3$ respectively. 
The $T$ is set to 0.005. The results are reported in Table \ref{table:phv}. Comparing to the Cycle-GAN, Coronary-GAN generates virtual histology images that are more similar to the real histology images, which is confirmed by multiple variants of PHV scores.
\subsubsection{Coronary-GAN reveals human coronary structure}
Two representative OCT images and their corresponding virtual histology images are shown in Figure \ref{fig:CoronaryGAN_result}. 
The virtual histology images generated by Coronary-GAN are capable of revealing the intima, media, adventitia layers, and their interfaces. The virtual histology images generated by Coronary-GAN is capable of providing visually similar images to the example histology image in Figure \ref{fig:CoronaryGAN_result}(a). On the contrary, the vanilla Cycle-GAN model fails to characterize layered structure in virtual histology. The performance of Cycle-GAN is substantially deteriorated when virtually staining the media and adventitia layers. Without incorporating structural information of human coronary, the Cycle-GAN can only use pixel intensities to perform virtual staining. The OCT signal attenuation over the sample depth will gradually decrease the pixel intensities, which explains the deteriorated performance of virtual staining media and adventitia layers that are deeper in depth. Moreover, the Cycle-GAN is prone to artifacts due to the lack of structural constrain. 
\begin{table}[t]
\centering
\caption{The PHV scores calculated using $i=1$ (PHV$\_1$), $i=2$ (PHV$\_2$), and $i=3$ (PHV$\_3$).}
\refstepcounter{table}
\label{table:phv}
\begin{tabular}{llll} 
\hline
\diagbox{Method}{Metric}         & PHV\_1 & PHV\_2 & PHV\_3  \\ 
\hline
\multicolumn{1}{c}{Cycle-GAN}    & 55.989 & 51.389 & 49.996  \\
\multicolumn{1}{c}{Coronary-GAN} & 61.561 & 61.308 & 56.714  \\ 
\hline
                                 &        &        &        
\end{tabular}
\vspace{-15pt}
\end{table}

%Moreover, the Cycle-GAN model is  signal decaying in depth as shown in Figure \ref{fig:CoronaryGAN}. signal decay in depth. For the media and adventitia that are deeper in depth, the Cycle-GAN 
\section{CONCLUSIONS}
We have developed a virtual staining network, namely Coronary-GAN, for human coronary samples using OCT images. Our Coronary-GAN is optimized for virtually staining human coronary OCT images by incorporating structural constrains. In future, we plan to improve the inference speed of Coronary-GAN, enabling real-time virtual staining. The proposed Coronary-GAN is promising for assisting the diagnosis of CAD by providing real-time, color-channel histology information using OCT images.

\section{COMPLIANCE WITH ETHICAL STANDARDS}
The human coronary autopsy specimens were de-identified and not considered as
human subjects, in compliance with the UAB at Birmingham’s Institutional Review Board (IRB).
\section{Acknowledgments}
This work was supported in part by National Science Foundation
(CRII-1948540), New Jersey Health Foundation, the National Center for
Advancing Translational Research of the National Institutes of Health
under award number UL1TR003096.
\label{sec:acknowledgments}
\bibliographystyle{IEEEbib}
\bibliography{strings,refs}

\end{document}